\RequirePackage{fix-cm}
\documentclass[twocolumn]{svjour3}          
\smartqed  
\usepackage{graphicx}  
\usepackage{dcolumn}   
\usepackage{bm}        
\usepackage{amssymb}   
\usepackage{amsmath}
\usepackage{xcolor}
\usepackage{float}
\journalname{Journal of Computational Electronics}

\begin{document}

\title{Tunable Electronic Properties of Multilayer Phosphorene and Its Nanoribbons
\thanks{The work was primarily supported by the NSF (DMR-1121288) through the UW-Madison MRSEC seed project. Supplementary funds were provided by the Splinter Professorship in Electrical and Computer Engineering. The work was performed using the resources of the UW-Madison Center for High Throughput Computing (CHTC).}
}


\author{S. Soleimanikahnoj and I. Knezevic}

\institute{The authors are with the
        \at Department of Electrical and Computer Engineering, University of Wisconsin-Madison, Madison, WI 53706, USA
              \email{soleimanikah@wisc.edu}; \email{iknezevic@wisc.edu}}

\date{Received: February 05, 2017 / Accepted: July 06, 2017}

\maketitle

\begin{abstract}
We study the effects of a vertical electric field on the electronic band structure and transport in multilayer phosphorene and its nanoribbons. {In phosphorene, at a critical value of the vertical electric field ($E_c$), the band gap closes and the band structure undergoes a massive-to-massless Dirac fermion transition along the armchair direction. This transition is observable in quantum Hall measurements, as the power-law dependence of the Landau-level energy on the magnetic field $B$ goes from $\sim (n+1/2)B$ below $E_c$, to $\sim [(n+1/2)B]^{2/3}$ at $E_c$, to $\sim [(n+1/2)B]^{1/2}$ above $E_c$. In multilayer phosphorene nanoribbons (PNRs), the vertical electric field can be employed to manipulate the midgap energy bands that are associated with edge states, thereby giving rise to new device functionalities.} We propose a dual-edge-gate PNR structure that works as a quantum switch.
\end{abstract}


\section{Introduction}
Black phosphorus (BP) is a thermodynamically stable allotrope of phosphorus with a layered structure. The layers of covalently bonded atoms are held together by the van der Waals interaction. Similar to obtaining graphene from graphite by mechanical exfoliation, BP can be isolated to a few layers~\cite{lu2014plasma,liu91acs}. The resulting structure is a recent addition to the family of two-dimensional (2D) materials called multilayer phosphorene. Monolayer phosphorene has a direct band gap of 1.45 eV~\cite{liu91acs}. In multilayer phosphorene, the gap decreases with increasing number of layers owing to relatively strong van der Waals interactions between the layers and remains direct ~\cite{zhang2014extraordinary}. This makes phosphorene a promising candidate for  electronic and optical applications~\cite{tran2014layer,jing2015small,buscema2014fast,koenig2014electric,xia2014rediscovering,ccakir2015significant,ccakir2014tuning}. Moreover, the phosphorene crystal structure is highly anisotropic, which gives rise to phenomena such as anisotropic electronic and thermal transport ~\cite{fei2014strain,fei2014enhanced,xia2014rediscovering,yuan2015transport,qin2015anisotropic,cai2015giant},  linear dichroism ~\cite{xia2014rediscovering,yuan2015transport}, and anisotropic plasmons ~\cite{Low2014Plasmons}. Additionally, owing to its heavily puckered structure, phosphorene is highly tunable by strain~\cite{elahi2015modulation,fei2014strain} and electric field~\cite{fei2014enhanced,das2014tunable}. In particular, applying an electric field normal to the layers reduces the band gap, leading to a transition from a moderate-gap semiconductor to a semimetal~\cite{kim2015observation}. Applying an electric field also leads to the emergence of more exotic features of phosphorene, including nontrivial topological phases~\cite{dutreix2016laser,liu2015switching,low2015topological}. However, the transition of the Landau levels (LLs) under the influence of electric field as a trademark of the topological phase transition has not been fully understood. {Furthermore, little is known about the possible practical use of electric-field modulation of the electronic characteristics of nanostructures based on phosphorene, such as nanoribbons \cite{carvalho2014phosphorene,Guo2014JPCC,Ali2015JPhysD,peng2014edge}.}

\begin{figure}
  \begin{center}
  \includegraphics[width=1\columnwidth]{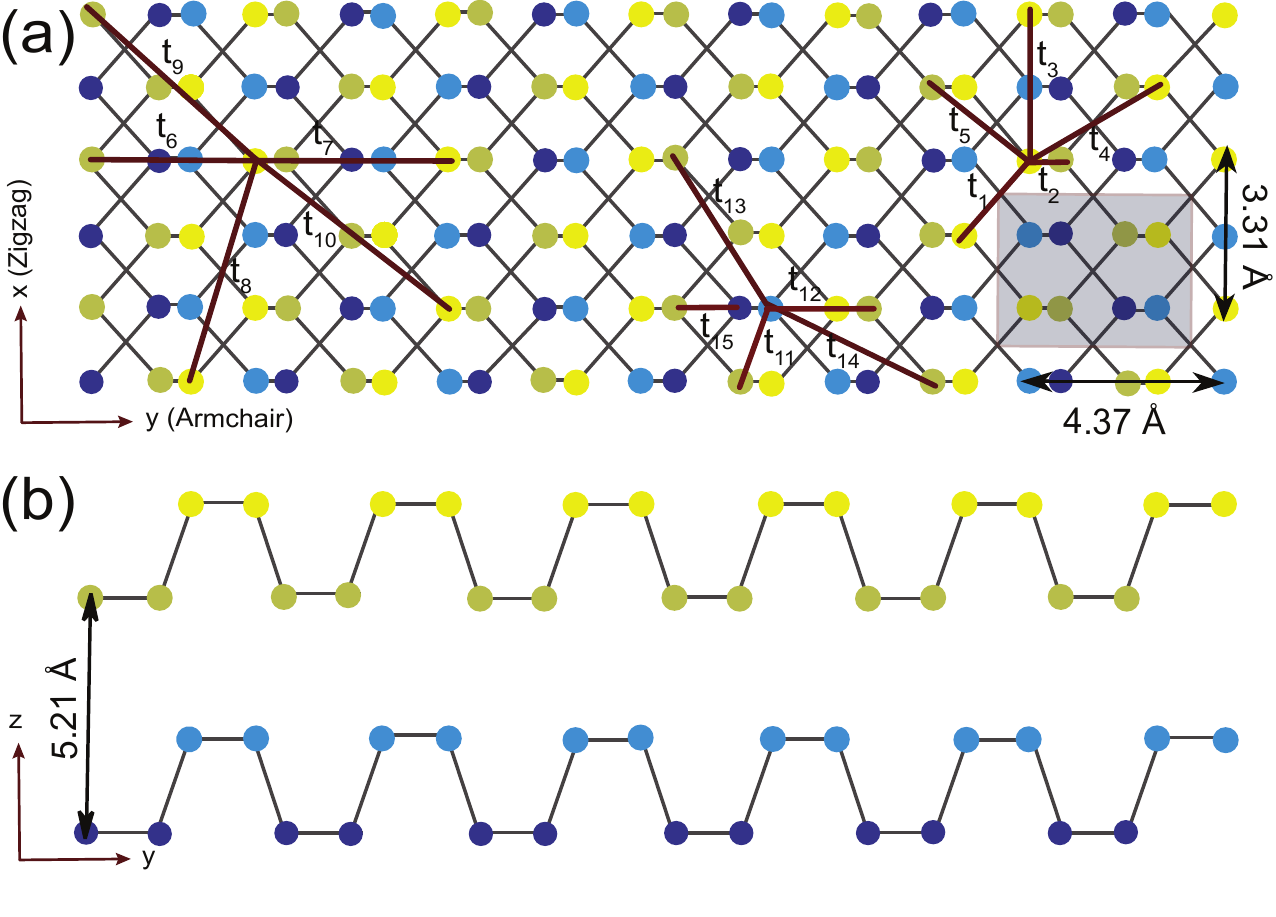}
  \end{center}
  \caption{\label{fig1} Schematic of bilayer phosphorene. (a) Top view. (b) side view. The left and the right edges are zigzag, while the top and bottom edges are armchair. Yellow (blue) circles are the phosphorene atoms in the upper (lower) layer while the atoms in the upper (lower) sublayer are portrayed a bright (dark) color. The gray rectangle denotes the unit cell of bilayer phosphorene. Higher multilayer systems will have the unit cell with the same top view and with additional 4 atoms per layer.}
\end{figure}

In this paper, we investigate the effects of a vertical electric field on the electronic properties (band structure and electronic transport) of multilayer phosphorene and its nanoribbons. In multilayer phosphorene at low fields, electrons with momenta in the zigzag direction [Fig. \ref{fig1}(a)] have parabolic bands, but in the armchair direction they behave as massive Dirac fermions with a gap-dependent effective mass. At a critical electric field $E_c$, the gap closes, and electrons exhibit a massive-to-massless Dirac fermion transition. Above $E_c$, there are two Dirac points, and the band structure is that of anisotropic massless Dirac fermions. This continuous massive-to-massless Dirac fermion transition could be observed in Hall measurements, as the LL energy dependence on the magnetic field would change from linear below $E_c$, to the novel 2/3-power at $E_c$, to the square-root dependence above $E_c$. {If two-dimensional (2D) bulk phosphorene is nanostructured in one dimension (1D), the electronic properties of the resulting quasi-1D phosphorene nanoribbon (PNR) will strongly depend on the direction of confinement and the type of edge termination. In particular, metallic multilayer PNRs have twofold-degenerate bands within the bulk gap and the associated wave functions are localized near the ribbon edges. We show that the application of a vertical electric field in metallic PNRs can selectively affect electrons in these midgap bands and thereby give rise to novel functionalities.} We propose a dual-edge-gate structure that affects these midgap states and drives the conducting-to-insulating transition in PNRs, thus enabling field-effect transistor action compatible with modern nanoelectronics, and potentially leading to new PNR-based devices.

\section{The Tight-Binding Model}
 The band structure of multilayer phosphorene is described by a tight-binding (TB) Hamiltonian
\begin{equation}
\label{eq1}
H = \sum_{i,j} t_{i,j}c^{\dagger}_{i}c_{j},
\end{equation}
which was parametrized on the basis of first-principles calculation within the $GW_0$ approximation~\cite{rudenko2015toward,rudenko2014quasiparticle}. As the TB parametrization was benchmarked for mono-, {bi-} and trilayers, we restrict our study here to these three systems, but note that TB may be suitable for larger multilayers, as well. A phosphorene monolayer contains two sublayers in a puckered structure and has four atoms per unit cell. Each additional layer brings four more atoms to the unit cell (see the structure of bilayer phosphorene in Fig. \ref{fig1}), with odd-numbered  layers aligned with other odd-numbered ones, and analogously for even-numbered layers. A unit cell for bilayer phosphorene is denoted by the shaded box in Fig. \ref{fig1}(a). In the nearest-neighbor approximation, there are fifteen relevant tight-binding hopping parameters (Table \ref{T1}), ten intralayer and five interlayer. The multilayer phosphorene structure is highly anisotropic. Cutting phosphorene in the (horizontal) $y$ direction (see Fig. \ref{fig1}) would result in an armchair edge, while cutting along the (vertical) $x$ direction would result in a zigzag edge. We refer to the $y$-direction as armchair and the $x$-direction as zigzag.

\setlength{\tabcolsep}{3.5pt}
\begin{table}[]

 \caption{Hopping parameters for the tight-binding Hamiltonian in Eq. (\ref{eq1}). $i=1-10$ are intralayer and $i=11-15$ are interlayer coupling terms.\label{T1} }
\begin{tabular}{ccccccccc}

 \hline
 \hline
 \vspace{.2cm}

$i$ & $t_{i } \mathrm{(eV)}$ & $d_{i} (\mathrm{\AA})$ & $i$ & $t_{i}  \mathrm{(eV)}$ & $d_{i} (\mathrm{\AA})$ & $i$ & $t_{i}  \mathrm{(eV)}$  & $d_{i} (\mathrm{\AA})$\\
  \hline
  1 & $-1.486$    &$ 2.22$  &6  & $+0.186 $ & $4.23$  & $11$  & $+0.524$ & $3.60$ \\
  2 & $+3.379$    &$ 2.24$  &7  & $-0.063 $ & $4.37$  & $12$  & $+0.180$ & $3.81$ \\
  3 & $-0.252$    &$ 3.31$  &8  & $+0.101 $ & $5.18$  & $13$  & $-0.123$ & $5.05$ \\
  4 & $ -0.071$   &$ 3.34$  &9  & $-0.042 $ & $5.37$  & $14$  & $-0.168$ & $5.08$ \\
  5 & $+0.019$    &$ 3.47$  &10 & $+0.073 $ & $5.49$  & $15$  & $+0.005$ & $5.44$ \\
\hline
 \hline
\end{tabular}
 \end{table}

\section{{ Bulk Multilayer Phosphorene in a Vertical Electric Field}} 
\begin{figure}
  \begin{center}
   \includegraphics[width=1\columnwidth]{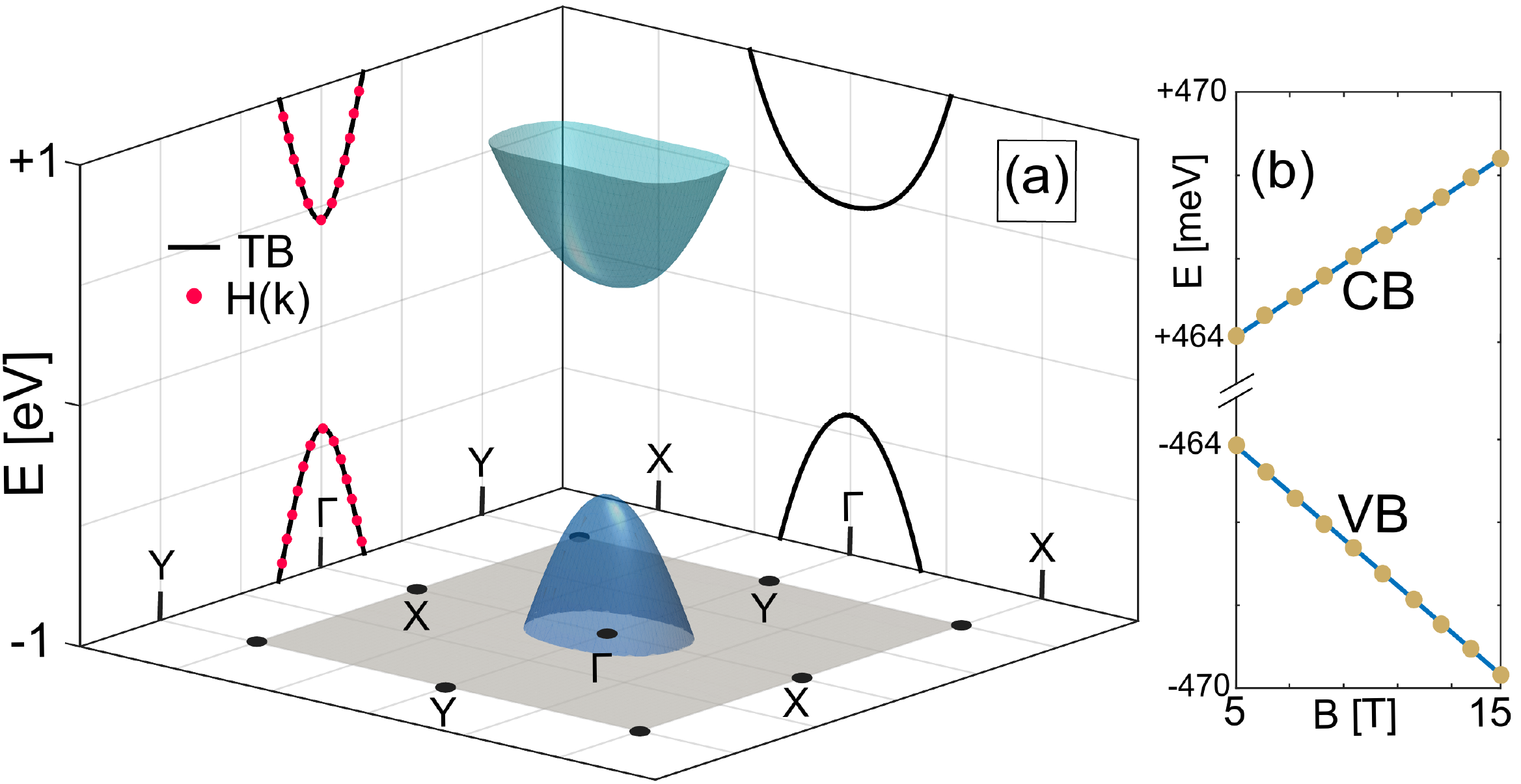}
  \end{center}
  \caption{\label{fig2} (a) band structure of unbiased trilayer phosphorene over the first Brillouin zone. Cuts normal to (10) and (01) through the Dirac point are projected onto the side walls, with the high-symmetry $\Gamma$, X, and Y points denoted. The solid lines are from TB, the red dotted line from the low-energy two-band Hamiltonian (\ref{foo}). { (b) The dependence of the conduction band (CB) and valence band (VB) edge (i.e., the lowest/highest Landau level in the CB/VB, respectively) on the external magnetic field for trilayer phosphorene in the absence of a vertical electric field. The blue lines are linear fits to the yellow dots obtained from Harper's equation.}}
\end{figure}

The band structure of trilayer phosphorene is depicted in Fig. \ref{fig2}(a); note the rectangular Brillouin zone (BZ) and its high-symmetry $X$ and $Y$ points along the $x$ and $y$ directions.  The band gap, $E_g$, is at the $\Gamma$ point and has a value of $0.85$ eV. Along the zigzag direction, both the conduction band (CB) and the valence band (VB) show quadratic dependencies on the wave vector $\bm{k}$ [Fig. \ref{fig2}(a)]. We calculate the effective masses of $m_c=3.18 m_{0}$ (CB) and $m_v=0.84 m_{0}$ (VB), where $m_0$ is the free-electron rest mass. In the armchair direction, however, the dispersion has an asymptotic linear trend. To describe this highly anisotropic band structure near the band gap, we propose a low-energy two-band Hamiltonian as
\begin{equation}
  \label{foo}
H(\bm{k}) = \bm{h}(\bm{k})\cdot\bm{\sigma},\,
    \bm{h}(\bm{k}) = \left[\frac{\hbar^2k_x^2}{2m_{c(v)}} + \frac{E_g}{2},\hbar v_{y}k_{y},0\right].
  \end{equation}
Here, $\bm{\sigma}$ are the Pauli matrices, $m_{c}$ and $m_{v}$ are the effective masses in the zigzag direction, and $v_{y} = 7.4\times 10^5  \mathrm{m/s}$ is the Fermi velocity in the armchair direction. In Fig. \ref{fig2}(b), we see that the fit from this effective two-band Hamiltonian (red dots) agrees very well with the CB and VB dispersions obtained by TB (solid curve) within $\pm 1$ eV of midgap. Finding the dispersion for $k_{x}=0$ from the two-band Hamiltonian (\ref{foo}) leads to
\begin{equation}
\label{eq3}
E(0,k_y) = \pm\sqrt{\left(\hbar k_{y} v_{y}\right)^2 + \left(\frac{E_{g}}{2}\right)^2}.
\end{equation}
This is a characteristic dispersion of massive relativistic particles ~\cite{weinberg1972gravitation}, where $E_{g}$ plays the role of rest energy. Upon employing the relativistic definition of the effective mass as $m_{y} = \hbar^2k \big(dE/dk_{y}\big)^{-1}$, we obtain an effective mass for both CB and VB in the armchair direction to be $m_y = E_g/2v_{y}^2 = 0.12m_{0}$, a value close to the experimentally obtained $0.08m_{0}$ ~\cite{kim2015observation}, which could not be explained previously based on the parabolic approximation. Thus, it is important to consider electrons in the armchair direction as massive Dirac fermions.

Since unbiased phosphorene is a conventional 2D electron gas, the CB and VB edges are expected to have linear dependencies on an applied vertical magnetic field~\cite{zhou2015landau,pereira2015landau,ghazaryan2015aspects}; { the CB/VB edge in a magnetic field is the lowest/highest Landau level (LL) in the CB/VB, respectively}. In order to test this hypothesis, magnetic field is incorporated in the tight-binding model through the Peierls substitution, which adds a phase to the hopping term between any two sites,

\begin{equation}
t_{i,j} \to  t_{i,j}e^{\frac{e}{h} \int_{R_i}^{R_j} \bm{A}.d\bm{l}},
\end{equation}
where $\bm{A} = (-By,0,0)$ is the vector potential and $B$ is the magnitude of the magnetic induction in $z$-direction. By this substitution, one will arrive at Harper's equation ~\cite{Harper1955,Thouless1982,Wakabayashi1999,Wiegeman1984}, and LLs are obtained by numerical calculation of its eigenvalues. For unbiased trilayer phosphorene, the LLs were calculated at various magnetic fields. The CB and VB edges are shown in Fig. \ref{fig2}(b). Predictably, the CB and VB edge change linearly as a function of magnetic flux density $B$.
{(While our calculations of LLs versus $B$ presented throughout this paper hold down to low magnetic fields, LLs are unlikely to be resolved in experiment at low fields owing to disorder ($B\mu\gg 1$ is needed to resolve LLs, where $\mu$ is the electron mobility in the sample). Therefore, we present the dependence of the lowest Landau levels in the conduction and valence bands on $B$ in the experimentally relevant high-magnetic-field range that was previously used by Jiang \textit{et al}. \cite{Jiang2007PRL} for the Hall measurement on graphene.)}

\begin{figure}
  \begin{center}
   \includegraphics[width=1\columnwidth]{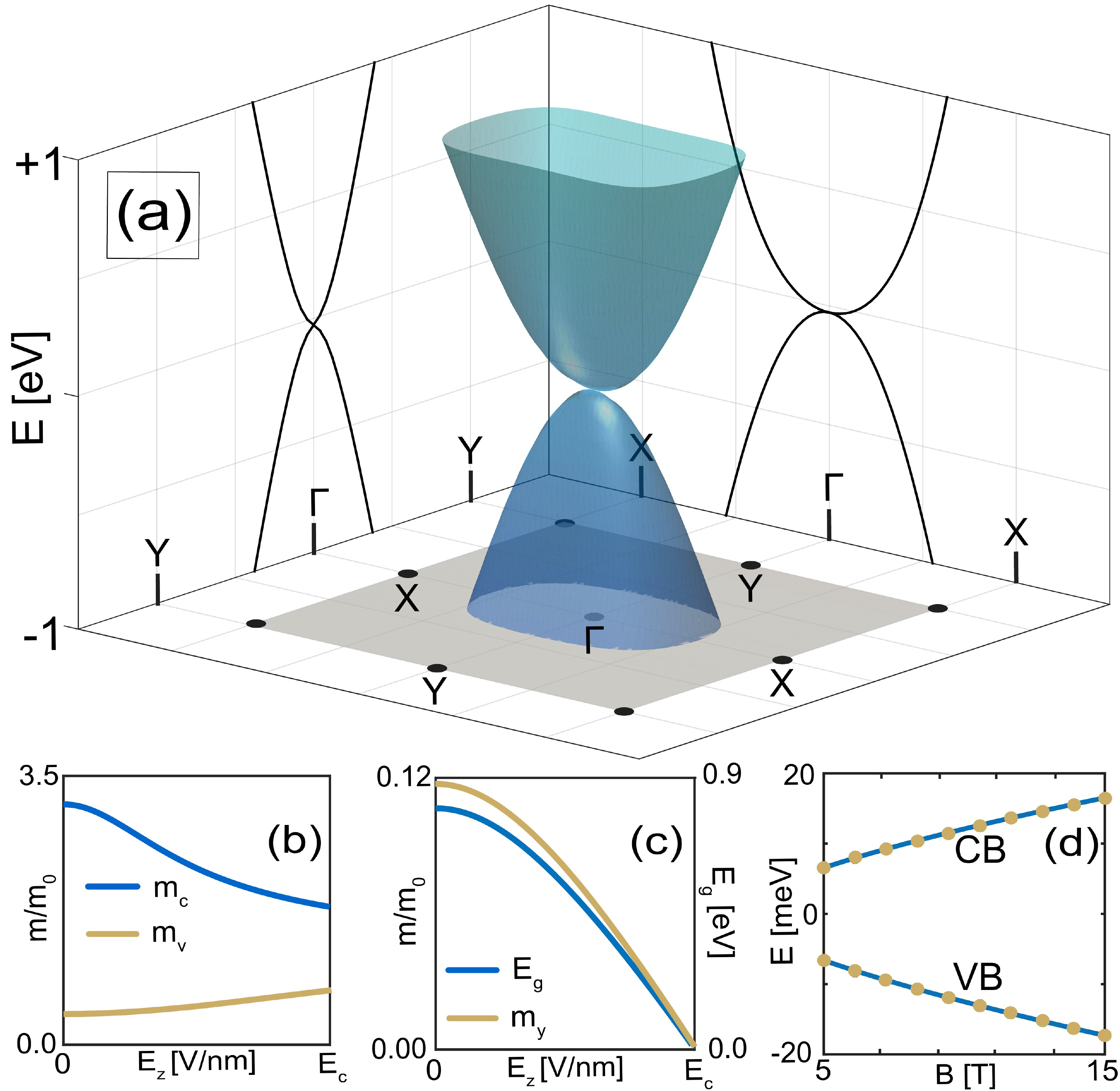}
  \end{center}
  \caption{\label{fig3} (a) band structure of biased trilayer phosphorene at the critical electric field ($E_z=E_c$). (b) Effective mass in the conduction and valence bands as a function of electric field along the zigzag direction. (c) Band gap and relativistic effective mass (both CB and VB) in the armchair direction versus electric field. (d) Energy of the lowest LL in the CB and VB versus magnetic field $B$, obtained by solving Harper's equation (dots). Solid lines are fits based on Eq. (\ref{eqLLEc}).}
\end{figure}

Next, we extend our model to the case of a vertical electric field $E_z$ [applied normal to the layers, i.e., in the $z$-direction in Fig. \ref{fig1}(a)]. In the TB calculations, $E_z$ is accounted for by assuming a linear potential drop across the structure extending from $-h/2$ to $h/2$ ($h$ is the thickness): a potential energy $V_i=eE_z(-h/2+z_i)$ is added to the diagonal terms of the TB Hamiltonian in Eq. (\ref{eq1}) according to \textit{i}-th atom's $z$ coordinate. Figure \ref{fig3}(c) shows that, as $E_z$ increases, the CB and VB shift toward each other due to the Stark effect. The band gap closes at a critical electric field, $E_c$,~\cite{kim2015observation} which decreases with increasing number of layers: $E_c\simeq0.17$ {V/\AA} for bilayer (also reported in~\cite{yuan2016quantum}) and $E_c\simeq0.15$ {V/\AA} for trilayer phosphorene. The full band structure of trilayer phosporene at $E_c$ is depicted in Fig. \ref{fig3}(a). Electric field affects the curvature of the parabolic CB and VB bands along the zigzag direction [Fig. \ref{fig3}(b)]: $m_{c}$ ($m_{v}$)  decreases  (increases) with increasing $E_z$, but $m_{v}<m_{c}$ for all values of $E_z$, in keeping with the high hole mobility reported in experiment~\cite{liu91acs}.

Application of the vertical electric field and the corresponding gap reduction have a strong effect on dispersion in the armchair direction: as $m_y \propto E_g$, $m_y$ drops with increasing $E_z$ and reaches zero at $E_{c}$ [Fig.\ref{fig3}(c)]. Therefore, there is a smooth transition from massive to massless Dirac fermions in the armchair direction under a vertical electric field. At $E_c$, the dispersion of CB and VB near the gap becomes
\begin{equation}\label{eq4}
E(\bm{k}) = \pm\sqrt{\left(\hbar k_{y} v_{y}\right)^2 + \left(\frac{\hbar^2k_x^2}{2m_{c(v)}}\right)^2}.
\end{equation}
Owing to the anisotropic dispersion at the critical electric field -- massless Dirac [Eq. (\ref{eq4})] along $y$, parabolic along $x$ -- the electron density of states (DOS) has a peculiar energy dependence: $\rho(E) \sim \frac{m}{v_y}\sqrt{E}$.~\cite{hasegawa2006zero} If such an electron system were placed in a uniform vertical magnetic field $B$, it would exhibit a unique dependence of the LL energy on the $B$ field (stemming from the square-root energy dependence of the DOS~\cite{LLCriticalFieldNote}) that has never before been observed in experiment:
\begin{equation}\label{eqLLEc}
E_n \sim \left[\left(n+ \frac{1}{2}\right)B\right]^{2/3},
\end{equation}
We obtained this dependence [see Fig. \ref{fig3}(d)] by numerically solving Harper's equation, as described earlier. The $2/3$-power dependence is distinct from the linear dispersion in two-dimensional electron gases, $E_n \sim \left(n + \frac{1}{2}\right)B$ [shown in Fig. \ref{fig2}(b)], and from the square-root dependence characteristic of Dirac fermions, $E_n \sim \sqrt{\left( n + \frac{1}{2}\right)B}$. Biased multilayer phosphorene would be the first realization of this LL dispersion in a real material, though the phenomenon was thoroughly investigated on a parametric honeycomb lattice~\cite{dietl2008new,montambaux2009merging,montambaux2009universal}. While Yaun \textit{et al.}~\cite{yuan2016quantum}, recently investigated LLs in perpendicular electric fields, they did not solve the Harper's equation~\cite{Harper1955,Thouless1982,Wakabayashi1999,Wiegeman1984} in an extended system, as we did here, and did not emphasize this peculiar behavior at $E_c$. We propose an extension to experiment~\cite{kim2015observation}, where the gap is closed via doping and mimics applying $E_c$: a measurement of the Hall conductance at this condition would experimentally confirm the peculiar form (\ref{eqLLEc}) of the LL $B$-field dependence.

\begin{figure}
   \begin{center}
    \includegraphics[width=\columnwidth]{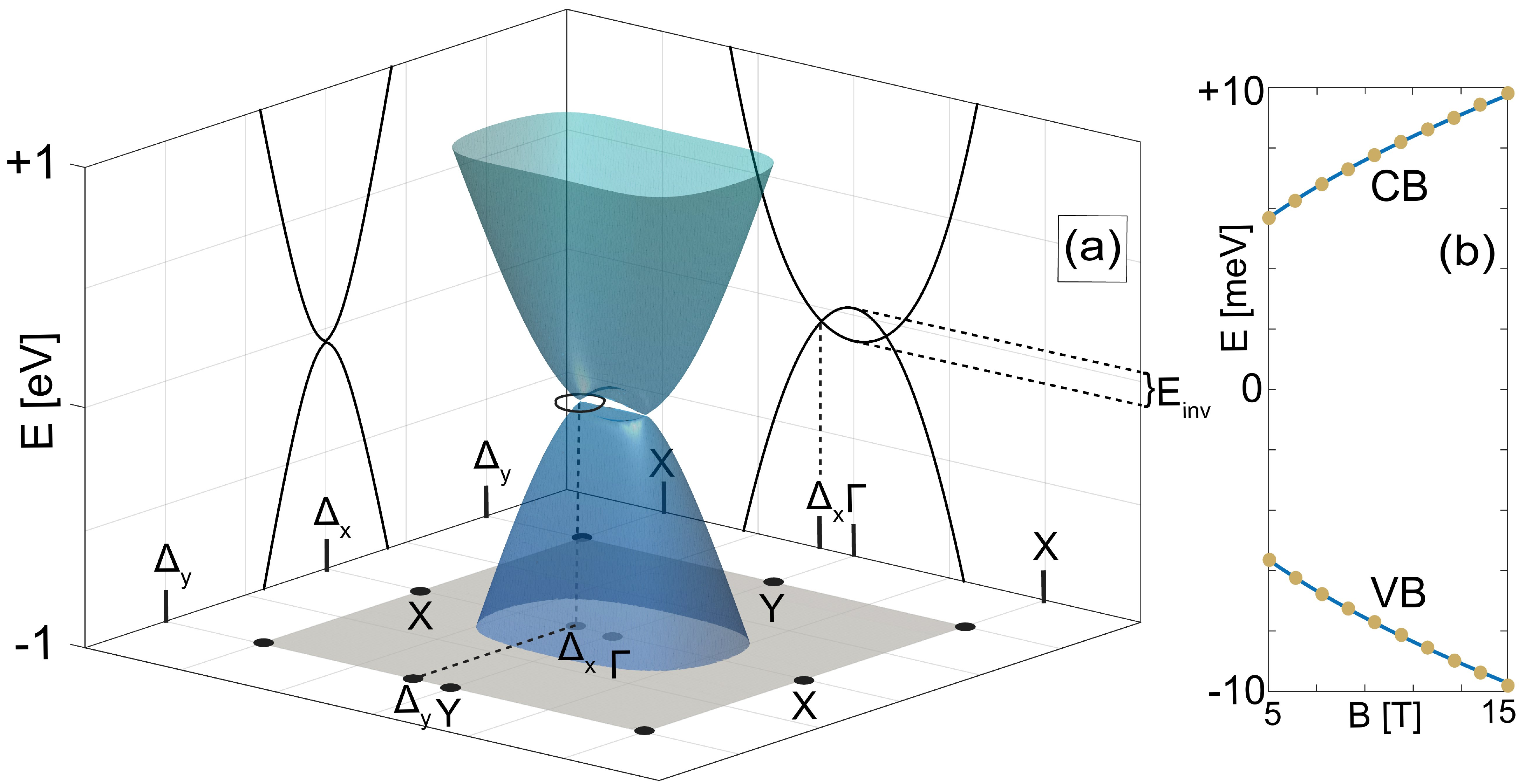}
   \end{center}
   \caption{\label{fig4} (a) band structure of biased trilayer phosphorene at $E_z >E_c$. A pair of Dirac points appear in the band structure ($\Delta_x$). (b) The CB and VB edge (yellow dots) as a function of magnetic field obtained from diagonalization of Harper's equation at $E_z > E_c$. The data is fitted with the square root power law of Dirac fermions (blue lines).}
 \end{figure}

Increasing the electric field beyond the critical value ($E_z > E_c$) splits the Dirac point into two. The Dirac points ($\Delta_X$) move away from the $\Gamma$ point along the X direction as a function of electric field [see Fig. \ref{fig4}(a)]. In the low-energy, two-band Hamiltonian (\ref{foo}), $\bm{h}(\bm{k})$ takes the form

\begin{equation}\label{eq6}
      \bm{h}(\bm{k}) = \left[\frac{\hbar^2k_x^2}{2m_{c(v)}} - \frac{E_{inv}}{2},\hbar v_{y}k_{y},0\right],
\end{equation}
where $E_{inv}$ is the value of the inverted gap shown in Fig. \ref{fig4}(a). At each Dirac point, the Hamiltonian can be expanded in terms of $\bm{q} = \bm{k} - \bm{k}_{\Delta_{X}}$, which gives us the
low-energy form $H(\bm{q}) = \hbar v_xq_x + \hbar v_yq_y$, a generic Dirac Hamiltonian with anisotropic Fermi velocity $v_{x,c(v)}=\sqrt{E_{inv}/m_{c(v)}}$ and ultimately the dispersion relation
\begin{equation}
E(\bm{q}) = \sqrt{\left(\hbar v_xq_x\right)^2 + \left(\hbar v_yq_y\right)^2}
\end{equation}
Based on this dispersion, the Berry phase, $\Lambda$, can be calculated by integrating the Berry potential, $\nabla \Lambda = \frac{\hbar^2v_xv_y}{2E(q)^2} \left[-q_y,q_x\right]$,  over a closed path around each Dirac point; the integration path chosen here is a circle around each Dirac point, depicted in Fig. \ref{fig4}(a). $\Lambda = \pm\pi$ is obtained from both the two-band low-energy and TB Hamiltonians.~\cite{park2011berry,resta2000manifestations} This means that electrons in biased multilayer phosphorene carry an extra degree of freedom called the pseudospin. Dirac fermions exhibit unconventional quantum Hall effect, where LLs vary with $B$ as $E_n \sim \sqrt{\left( n + \frac{1}{2}\right)B}$ ~\cite{zhang2005experimental,yuan2016quantum}. The magnetic-field dependencies of the CB and VB edge of multilayer phosphorene biased at $E_z > E_c$ are shown in Figs. \ref{fig4}(b). As can be seen, both CB and VB (yellow circles) follow the square root power law of Dirac fermions (blue lines).

\section{{Metallic Phosphorene Nanoribbons in a Vertical Electric Field. Dual-Edge-Gate Device}}

\begin{figure}
  \begin{center}
   \includegraphics[width=1\columnwidth]{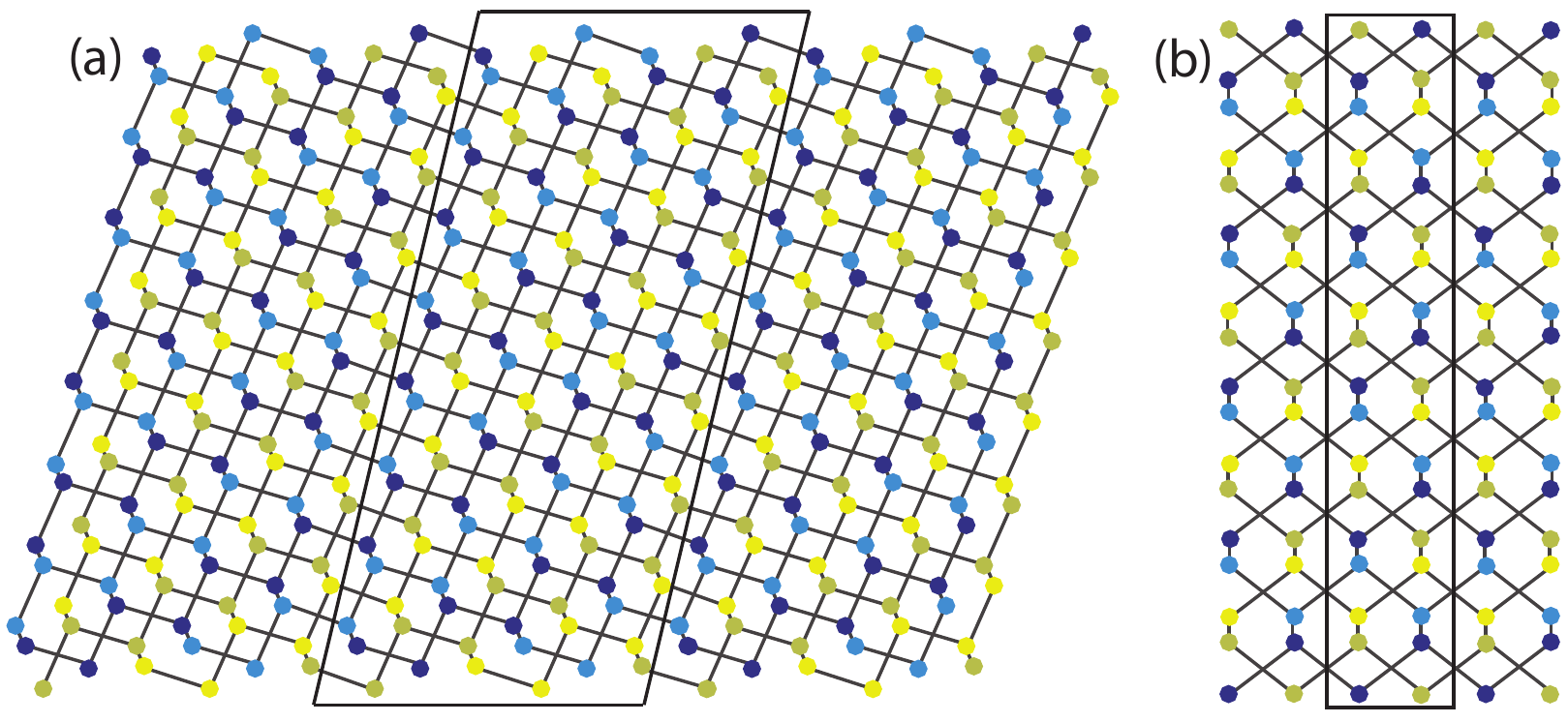}
  \end{center}
  \caption{\label{fig5} Crystal structure of bilayer skewed-armchair (a) and zigzag (b) PNRs. Corresponding unit cells are outlined by the solid line boxes.}
\end{figure}

{ Nanoribbons are quasi-1D systems with a high length-to-width ratio, often fashioned from the ultrathin quasi-2D materials. The nanoribbon geometry is more amenable to device applications than the quasi-2D sheet geometry, and the electronic properties of ultranarrow nanoribbons are very sensitive to edge termination. Phosphorene nanoribbons, in particular, can combine the effects of band structure anisotropy and sensitivity to external fields, which characterize the underlying 2D phosphorene, with the effects of edges to give rise to unique electronic properties. Indeed, armchair and zigzag PNRs nanoribbons have been attracting a lot of attention in recent years \cite{carvalho2014phosphorene,Guo2014JPCC,peng2014edge,Ramasubramaniam2014abinitio,Ali2015JPhysD}, with skewed-armchair nanoribbons having come into focus more recently \cite{grujic2016tunable}. Here, we are specifically interested in metallic PNRs, which are found among skewed-armchair and zigzag PNRs [see Figs. \ref{fig5}(a) and \ref{fig5}(b), respectively]. These metallic PNRs have interesting behavior in a vertical electric field, which, as we will show, gives rise to new potential device functionalities. Other types of PNRs are insulating and will not be considered in this paper ~\cite{grujic2016tunable,sisakht2015scaling,carvalho2014phosphorene}.}

\begin{figure}
  \begin{center}
   \includegraphics[width=\columnwidth]{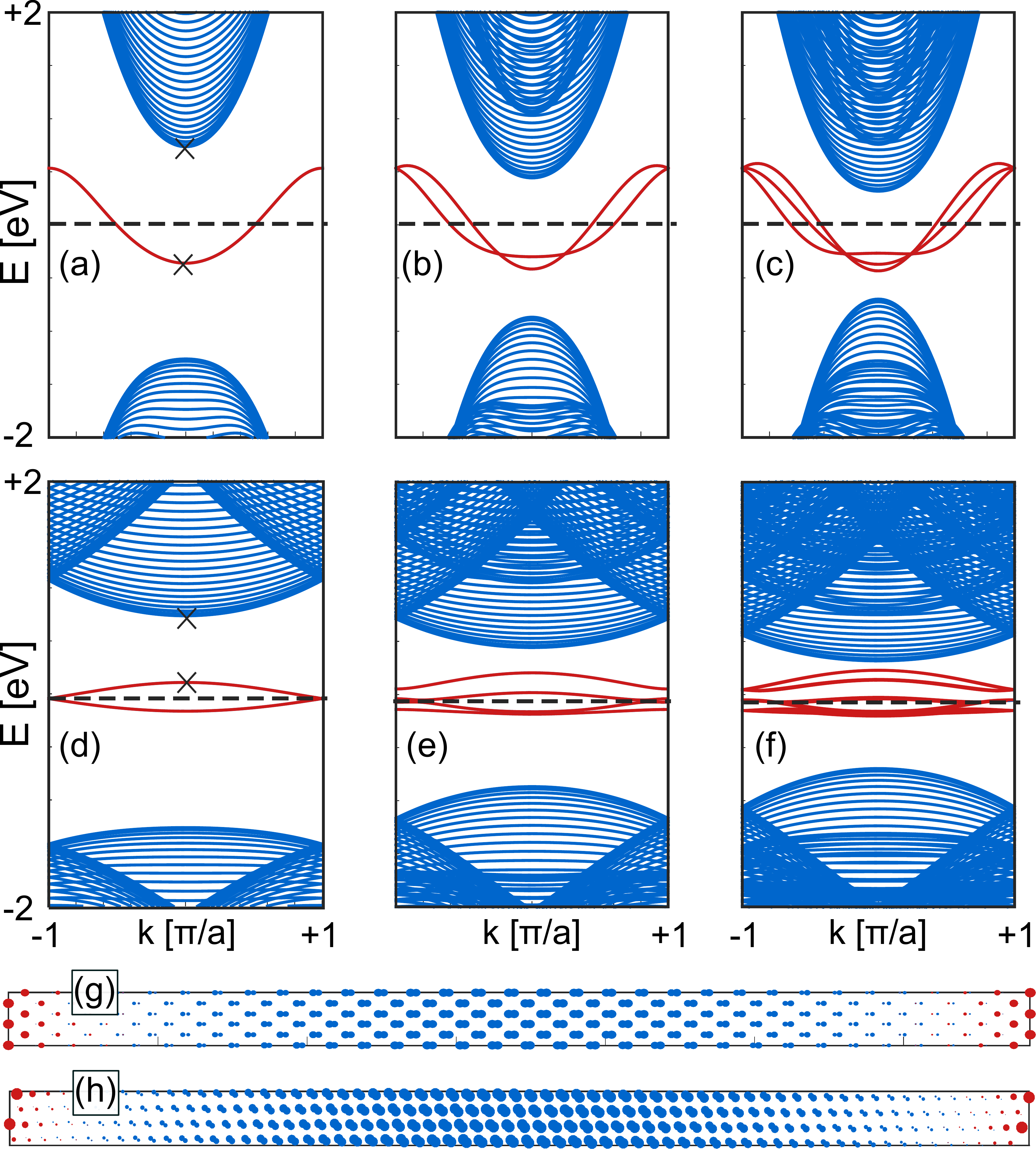}
  \end{center}
  \caption{\label{fig6} band structure of (a) monolayer, (b) bilayer, and (c) trilayer zigzag PNRs. band structures of their skewed-armchair counterparts are shown in (d)--(f), respectively. All ribbons are $18$ nm wide. (g) and (h) Probability densities of the states whose energies are marked by ``x'' in panels (a) and (d), respectively. The red (blue) circles denote the probability densities of the midgap (bulk) states shown by the same color in panels (a) and (d).}
\end{figure}

The calculated band structures of mono-, bi-, and trilayer zigzag PNRs are shown in Figs. \ref{fig6}(a)--(c) and those corresponding to skewed-armchair PNRs are shown in Figs. \ref{fig6}(d)--\ref{fig6}(f), respectively. One can see the presence of  midgap  bands (red curves) completely detached from the bulk bands (shown in blue). For zigzag PNRs, each layer of phosphorene contributes one band of twofold-degenerate midgap states, making a total of two, four, six midgap states for mono-, bi- and trilayer PNRs, respectively. However, in the case of skewed-armchair PNRs, each layer provides two, twofold-degenerate midgap states. As a result, a total of four, eight, twelve midgap bands are present for mono-, bi- and trilayer skewed-armchair PNRs, respectively. {We used the same tight-binding parameters for the PNRs as in the bulk calculation (Table 1). Of course, there is generally no guarantee that the bulk tight-binding parameters remain suitable for nanoribbons. However, there are at present no published DFT calculations for skewed PNRs against which we could benchmark tight-binding parameters. For ZPNRs, DFT calculations predict the existence of midgap bands \cite{carvalho2014phosphorene}, as does tight binding. For ultranarrow ZPNRs (width below 3 nm), the dispersions of midgap bands found using DFT \cite{Guo2014JPCC} differ somewhat from those predicted by the tight-binding model with bulk parameters \cite{sisakht2015scaling}, but as the width of the ribbons increases and the gap between  the bulk bands decreases, the results from the TB model become more consistent with the dispersions predicted by DFT \cite{Ramasubramaniam2014abinitio}.}

In the absence of a vertical electric field, zigzag PNRs are metallic since the Fermi level passes through all midgap bands and is far from the bulk states. In a similar manner, metallicity of skewed-armchair PNRs is caused by the two midgap states closest to the Fermi level without any contribution from the bulk states [see Figs. \ref{fig6}(d)-\ref{fig6}(f)]. Therefore, near-equilibrium electronic transport in metallic PNRs is governed by midgap states. In Figs. \ref{fig6}(g) and \ref{fig6}(h), we plot the probability density for two $k=0$ wave functions marked in Figs. \ref{fig6}(a) and \ref{fig6}(d), respectively, one from a midgap band (shown in red) and another from the bulk CB (shown in blue). In contrast to the states from the bulk bands, whose probability density peaks near the PNR middle, the midgap state's probability density is highest by the edges (near one edge in the top sublayer and near the other end in the bottom sublayer of a given monolayer). Therefore, electronic transport in PNRs should be tunable by the manipulation of the midgap states, confined to the edges.

\begin{figure}
  \begin{center}
   \includegraphics[width=.95\columnwidth]{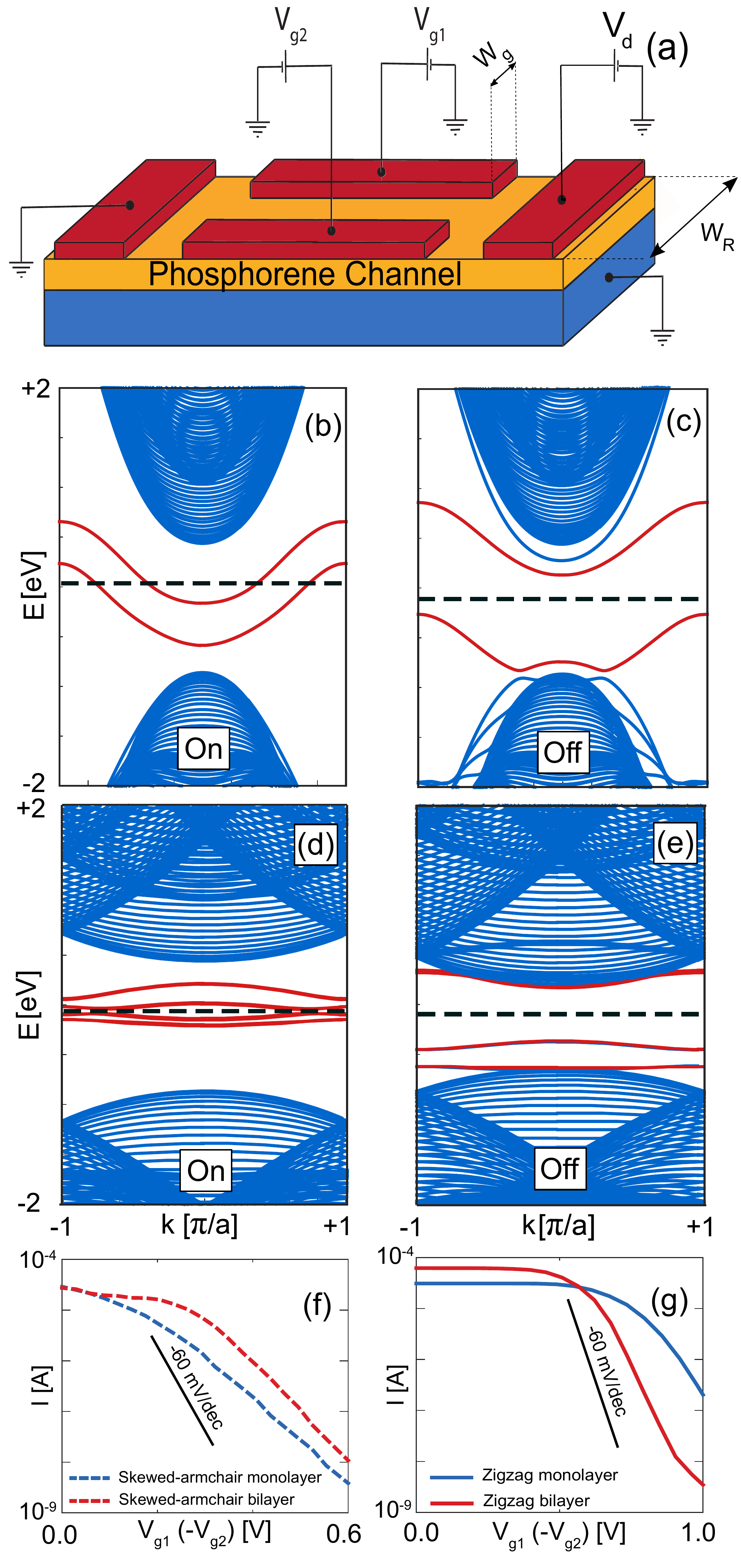}
  \end{center}
  \caption{\label{fig7}  (a) Schematic of the proposed dual-edge-gate field-effect structure to control the conductance in PNRs. (b) \& (c) [(d) \& (f)] band structure of zigzag (skewed-armchair) bilayer PNRs under applied bias on the edge gates. Plots in (b) and (d) are for bias $V_{g1} = -V_{g2} = +0.4$ V, while in (c) and (d) $V_{g1}=-V_{g2}=+0.8$ V. Red curves represent the dispersions of the twofold degenerate midgap bands, whose wave functions are confined near the PNR edges. The dashed horizontal line represents the Fermi level. {(f) and (g) Current in skewed-armchair (f) and zigzag (g) single and bilayer PNRs as a function of applied bias on the edge gates, indicating a transition from a conductor to an insulator. The results are limited to single and bilayer PNRs since in trilayer PNRs the van der Waals interaction between layers reduces the energy difference between the midgap states and the bulk states. As a result, bulk states are also affected by the bias of the gates and a transition to the ``off'' state is not accessible
 at room temperature in trilayer PNRs. The absolute values of the subthreshold swing are 136 and 112 mV/decade for single and bilayer skewed-armchair PNRs in (f), respectively, and 115 and 80 mV/decade for single and bilayer zigzag PNRs in (g), respectively. For reference, the turn-off corresponding to the 60 mV/decade subthreshold swing (theoretical limit in silicon MOSFETs) is also depicted.}}
\end{figure}

Based on this finding, we propose a field-effect transistor (FET) shown in Fig. \ref{fig7}(a), where the conductance in PNRs via the midgap states (which are confined near the edges) is modulated locally by two edge gates, with voltages $V_{g1}$ and $V_{g2}$. The width of the ribbons ($W_R$) is chosen to be $18$ nm, wide enough for experimental realization yet narrow enough to be computationally feasible via atomistic TB and nonequilibrium Green's functions (NEGF) ~\cite{datta1997electronic}. The width of each edge gate ($W_g$) is $5$ nm and the drain voltage ($V_d$) is fixed at $V_d = 50$ mV. Figure \ref{fig7}(b) and (d) shows the band dispersion for zigzag and skewed armchair PNRs for $V_{g1} = -V_{g2} = +0.4$ V respectively. The applied bias strongly affects the midgap states, which have a high probability density right under the gate. The field induced by the bias shifts the bands whose wave function is localized in the upper sublayer upwards and shift the bands whose wave function is in the lower sublayer downward.

These edge gates have essentially no effect on the bulk bands. Although the midgap states have moved toward the bulk bands compared to their initial position in the gap [see Figs. \ref{fig6}(b) and \ref{fig6}(e)], the ribbon  remains metallic since the Fermi level [which is fixed by the potentials on the source and drain on the left and right, Fig. \ref{fig7}(a)] still passes through the midgap bands. Increasing the bias voltages to $V_{g1}=-V_{g2}=+0.8$ V in Figs. \ref{fig7}(c) and \ref{fig7}(e) leads to a further energy separation of the midgap bands, so there are no states at the Fermi level, which denotes a transition from conducting [Fig. \ref{fig6}(b) and (d)] to insulating behavior [Fig. \ref{fig6}(c) and  \ref{fig6}(e)].

This transition is also demonstrated by calculating the current for single and bilayer PNRs at room temperature ($T=300\,\mathrm{K}$)) {[Figs. \ref{fig7}(f) and (g) for skewed-armchair and zigzag PNRs]} using the ballistic NEGF formalism.~\cite{datta1997electronic}. { We implemented the low-field ballistic NEGF transport technique without self-consistent electrostatics in a fairly standard way: equilibrium Green's functions are calculated at a given energy for the open system PNR system, where the role of contacts is incorporated in the NEGF via contact self-energies using the Sancho-Rubio iterative scheme \cite{SanchoRubio85}. The calculation of Green's functions is performed for a dense array of energies within the transport window, i.e., over the energy range between the source and drain Fermi levels (50 meV) plus/minus several $k_BT$ around this range ($k_B$ is the Boltzmann constant). The edge gates  are at potentials $V_{g1}$ and $V_{g2}$ with respect to the body potential, which is assumed zero, (i.e., body is grounded). The influence of the edge gates is incorporated into the relevant diagonal terms in the tight-binding Hamiltonian. It is assumed that the potential drops linearly over the PNR thickness, from the values of $V_{g1}$ and $V_{g2}$ at the top, underneath the gates, to zero at the bottom; in other words, the vertical field is constant under the gates. The points at the top of the PNR that are not under a gate are assigned zero potential. Finally, in the tight-binding NEGF transport calculation, the PNRs as implemented have lengths and widths in the several-nanometer range, but the actual lengths do not matter in the purely ballistic limit, as it is assumed that the edge gates go along the whole length of the PNR.}

At low bias on the edge gates, the current $I$ [Fig. \ref{fig7}(f)] is relatively high, indicating that midgap states are crossing the Fermi level or located at the vicinity of Fermi level. As the voltage magnitude increases, the midgap states are pushed apart. Above a threshold bias on the edge gates, the current becomes vanishingly small, demonstrating a transition to an insulator [Fig. \ref{fig7}(f)]. This dual-edge-gate structure is compatible with current fabrication technology, as similar multigate structures were realized previously ~\cite{craciun2009trilayer,williams2011gate}. { For single and bilayer zigzag PNRs [Fig. \ref{fig7}(g)], the subthreshold swing (SS) values are 115 and 80 mV/decade,  respectively. In case of single layer and bilayer skewed-armchair PNRs [Fig. \ref{fig7}(f)], the SS values are 136 and 112 mV/decade, respectively.  For reference, we also added the 60 mV/decade SS  curves to Fig. 7(f) and (g), which would correspond with the turn-off theoretical limit in Si MOSFETs.  The values obtained are lower than the 5 V/decade obtained for bulk-phosphorene  field-effect transistors \cite{Li2014NatureNano}, they are still higher than the SS in commercial silicon-based devices (70 mV/decade).}

A recent calculation showed that edge functionalization can lead to a dramatic flattening of the midgap-band dispersions ~\cite{peng2014edge}, which would significantly reduce the threshold voltage required for the conductor-to-insulator transition. { As the current in the on-state is channeled by the midgap states, the SS will be dependent on the dispersion of midgap states. Edge passivation of PNRs leads to almost dispersion-free midgap states \cite{peng2014edge}, and this effect could approximately be captured in the tight-binding model by setting $t_3 = 0$ \cite{ezawa2014topological}. Indeed, our preliminary calculations with $t_3 = 0$ yield a SS of 15 mV/decade, which is below the theoretical limit for ideal MOSFETs \cite{Choi2007IEEE_EDL}. This result hints that edge-passivated PNRs might be promising as field-effect switches.}

\section{Conclusion}
In summary, we showed that the electron band dispersion for multilayer phosphorene transitions from parabolic in the zigzag and massive Dirac fermion in the armchair direction at vertical electric fields below $E_c$ to a dispersion of an anisotropic massless Dirac fermion at electric fields above $E_c$. We posited that this transition would be observed in quantum Hall experiments as an electric-field-dependent change in the Landau-level energy vs $B$: $\sim (n+1/2)B$ below $E_c$, $\sim [(n+1/2)B]^{2/3}$ at $E_c$, and $\sim [(n+1/2)B]^{1/2}$ above $E_c$.

{Moreover, we showed that selective tuning of midgap bands in metallic PNRs results in novel device functionalities. We proposed a structure with dual edge gates that strongly affect the edge states associated with the midgap bands, but leave the bulk ones inert. The dual-edge-gate structure can induce a transition from a conducting ``on'' state to an insulating ``off'' state by moving the midgap bands away from the Fermi level, thereby realizing field-effect transistor action on PNRs.}

Electric-field modulation of phosphorene and PNRs is a versatile concept that can enable access to new physics and new functions in phosphorene-based devices.

\section{Acknowledgement}
The work was primarily supported by the NSF (DMR-1121288) through the University of Wisconsin Materials Research Science and Engineering Center (MRSEC) seed project. After the end of the NSF seed support, funding by the University of Wisconsin -- Madison through the Patricia and Michael Splinter Professorship in Electrical and Computer Engineering (awarded to I.K.) enabled successful completion of this work. The work was performed using the resources of the UW-Madison Center for High Throughput Computing (CHTC).



\end{document}